\documentclass[manuscript,screen]{acmart}
\usepackage{graphicx}
\usepackage{subcaption}

\AtBeginDocument{%
  }

\setcopyright{acmlicensed}
\copyrightyear{2024}
\acmYear{2024}
\acmDOI{XXXXXXX.XXXXXXX}

\acmConference[Accepted ACM SIGITE]{ACM SIGITE 2024}{October 10--12,
  2024}{El Paso, TX}
\begin{document}

\title{Faculty Perspectives on the Potential of RAG in Computer Science Higher Education}

\author{Sagnik Dakshit}
\email{sdakshit@uttyler.edu}
\orcid{0000-0001-9339-6259}
\affiliation{%
  \institution{The University of Texas at Tyler}
  \city{Tyler}
  \state{Texas}
  \country{USA}
}

\renewcommand{\shortauthors}{S. Dakshit}

\begin{abstract}
  The emergence of Large Language Models (LLMs) has significantly impacted the field of Natural Language Processing and has transformed conversational tasks across various domains because of their widespread integration in applications and public access. The discussion surrounding the application of LLMs in education has raised ethical concerns, particularly concerning plagiarism and policy compliance. Despite the prowess of LLMs in conversational tasks, the limitations of reliability and hallucinations exacerbate the need to guardrail conversations, motivating our investigation of RAG in computer science higher education. We developed Retrieval Augmented Generation (RAG) applications for the two tasks of virtual teaching assistants and teaching aids. In our study, we collected the ratings and opinions of faculty members in undergraduate and graduate computer science university courses at various levels, using our personalized RAG systems for each course.  This study is the first to gather faculty feedback on the application of LLM-based RAG in education. The investigation revealed that while faculty members acknowledge the potential of RAG systems as virtual teaching assistants and teaching aids, certain barriers and features are suggested for their full-scale deployment. These findings contribute to the ongoing discussion on the integration of advanced language models in educational settings, highlighting the need for careful consideration of ethical implications and the development of appropriate safeguards to ensure responsible and effective implementation.
\end{abstract}




\received{20 February 2007}
\received[revised]{12 March 2009}
\received[accepted]{5 June 2009}

\maketitle

\section{Introduction}
\label{Intro}

Deep learning (DL), a subset of Artificial Intelligence (AI), has greatly impacted the industry by allowing the extraction of meaningful patterns from real-world data, resulting in widespread applications in numerous domains and tasks, including education. The integration of deep learning into educational settings offers numerous opportunities to improve teaching and learning processes, including personalizing instruction, optimizing learning environments, and facilitating the development of innovative educational technologies. With the ability to automatically learn complex patterns and representations from large amounts of data, deep learning holds tremendous potential for enhancing teaching, learning, and educational outcomes. One recent disruptive innovation with the potential to significantly impact education is that of Large Language Models (LLMs), which have demonstrated significant potential in conversational natural language processing (NLP) tasks.

Large language models have been instrumental in revolutionizing the field of natural language processing, driving remarkable advancements in text generation, comprehension, and interpretation. By leveraging extensive datasets and advanced deep-learning architectures, particularly transformer models, LLMs can produce human-like text and handle a diverse array of tasks. From answering complex questions and summarizing lengthy documents to participating in meaningful dialogues, LLMs demonstrate an unprecedented ability to understand and generate natural languages. This transformative capability has not only improved automated communication systems but has also facilitated significant progress in applications such as machine translation, sentiment analysis, content creation, and conversational agents. The continuous evolution of LLMs and their applications promises greater potential for innovation and efficiency across various domains. Retrieval Augmented Generation (RAG) is an emerging application paradigm that seeks to combine the strengths of large-language models (LLMs) with external knowledge sources.  LLMs, which are powerful for generating coherent and contextually relevant text, face challenges \cite{kaddour2023challenges} in providing accurate, up-to-date, and domain-specific information because of their reliance on static training data. RAG addresses these limitations through a retrieval mechanism that enables the model to access external databases or knowledge sources at inference time, thereby enhancing its ability to produce more precise and informative responses. This integration of retrieval and generation opens new possibilities for applications such as question answering, summarization, and dialogue systems, allowing for a more dynamic and responsive approach to content generation that can pave a new path in education. 

In this study, we aim to provide educators, researchers, policymakers, and technology developers with insights into the potential and limitations of LLMs to reshape higher computer science education and foster innovation in teaching and learning. We developed specialized RAG models for faculty members and considered their feedback and opinions on identifying its use cases, potentials, and limitations. Through our discussions on the potential and limitations of LLMs in higher computer science education, we hope to accelerate the progress in digital innovation in higher computer science education that empowers learners and educators in the digital age. We obtained and reported faculty ratings and perspectives on two tasks of virtual teaching assistants and teaching aids as applications in higher educational settings. To the best of our knowledge, this is the first work that conducts deployed application-based studies and reports faculty opinions and ratings on the use of LLM-based RAG systems in computer science higher education discuss the potential and limitations in the use of LLM-based RAG in computer science higher education.

\subsection{Contribution}
The contributions of this study can be summarized as follows:
\begin{itemize}
    \item We developed a large language model-based Retrieval Augmented Generation system for computer science higher education Junior, Senior and Graduate level classes.
    \item We investigated the potential of RAG to serve as a virtual teaching assistant and as a teaching aid to faculty members in higher education.
    \item We discuss the limitations and improvements required for the integration of RAG for the advancement of digital education based on the collected faculty feedback and ratings.
\end{itemize}

The rest of the paper is structured with an illustration of our methodology in Section \ref{methodology} followed by a presentation of the collected faculty opinions in Section \ref{perspectives}. Finally, we discuss the potential improvements required for large-scale deployment and our future work in Section \ref{discussions}.

\section{Related Works}

In this section, we discuss relevant research on the application of LLM in general, and specifically for improving higher education. Owing to the nascent widespread development of LLMs and lack of widespread conversational education data, their applications and implications in education have been explored with limited capacity. Dempere et. al \cite{dempere2023impact} presented a comprehensive survey on the impact of AI chatbots like ChatGPT on Higher Education Institutes (HEIs) by employing a scoping review of the current literature. Most reported studies do not consider faculty opinions, nor do they investigate RAG systems for specific tasks. The predominant discussion has been on the ethics \cite{mvondo2023exploring} of LLM integration in education from the perspective of plagiarism, and it has also been suggested to go to old school pens and papers over digital methods \cite{milano2023large, mvondo2023generative}. In \cite{fowler2023first}, all applications and ethics of ChatGPT across all Australian universities over the first 100 days of its release were thoroughly discussed. A perspective study
on the potential of ChatGPT to facilitate adaptive learning, provide personalised feedback, support research and data analysis, offer automated administrative services, and aid in developing
innovative assessments are presented in \cite{rasul2023role}. While the above studies mostly explore the challenges and ethical implementations of GPT in higher education, \cite{tajik2024comprehensive} comprehensively examined the works on potential from student-facing, teacher-facing, and system-facing tools perspectives. These studies mostly focus on the limitations, applications, and ethics from a university policy-making view and do not pertain to any specific educational programs where the role and applications can be different. Moreover, no user study has been conducted, and the potential of LLM as a virtual assistant as a student and teaching aid has not been explored. 

In the context of computer science higher education, mitigation of plagiarism \cite{orenstrakh2023detecting}, the possibility of generating class lectures using LLM has been explored in \cite{christensen2023llm}, and learner performance exploration \cite{kumar2023impact} have been explored. In \cite{agarwal2024llm}, a non-exhaustive comparison of some LLMs was investigated for use in education but not from the perspective of virtual teaching assistants or teaching aids. The authors in \cite{arora2024analyzing} explored patterns of GPT usage by students over a semester-long computer science class. The closest study to our approach was conducted by Lu et al.. al \cite{lyu2024evaluating}, also conducted a semester long survey and collected conversational and message-level feedback rating from students on five prongs namely: 1) comprehension, 2) critical thinking, 3) syntax mastery, 4) independent learning, and 5) TA replacement likelihood. Their study demonstrated a shift towards human assistant support over virtual assistants. In contrast to their investigation of the opinions of faculty members and, unlike their use of OpenAI's GPT, we explore the potential of Retrieval Augmented Generation (RAG), which has the added benefit of curbing hallucinations and limiting responses to faculty-approved answers.

Authors Liu et. al explored the use of RAG in popular CS50 class \cite{liu2024teaching} and collected feedback from students verifying that “guardrailed” approach of RAG is superior over prohibiting the use of LLMs. However, no perspective or feedback from the faculty members was collected to identify the potential or limitations of this study. Furthermore, their task did not explore the specific point of teaching aid to faculty members or teaching assistants. Another notable study evaluated and graded students’ open-ended written answers using RAG models \cite{jauhiainen2024evaluating}.  In contrast to the above studies, our contributions focus on faculty feedback and opinions to understand the potential and limitations of RAG in their use as virtual teaching aids and virtual teaching assistants.

\begin{figure*}[h]
    \centering
    \includegraphics[width=5in,height=2.5in]{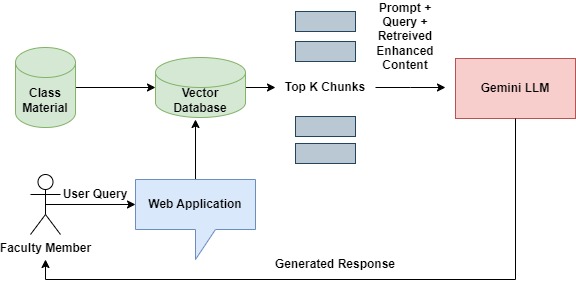}
    \caption{Our individual RAG pipeline using Gemini LLM developed on class materials provided by faculty members.}
    \label{fig:pipe}
\end{figure*}

\section{ Methodology}
\label{methodology}
\subsection{Research Design}
\label{design}
In this section, we discuss our experimental setup to investigate the potential and challenges of using LLM-based RAG in computer science higher education from the perspective of faculty members in a university setting. We designed our experiments to evaluate the feasibility of using LLM-based RAG as a teaching assistant and teaching aid. For our purposes, we defined the \emph{Teaching Assistant} as an aid to faculty members in answering questions commonly asked by students. In comparison, \emph{Teaching Aid} is defined as aid to faculties in the generation of study guides, quizlets, and assignment or exam questions. We did not investigate the students’ perspectives of virtual RAG teaching assistants in this study. Through this study, we attempted to answer three research questions:
\begin{itemize}
    \item RQ1: Can RAG be used as \emph{Teaching Aid} for faculty in computer science higher education.
    \item RQ2: Can RAG be used as \emph{Virtual Teaching Assistant} for students in computer science higher education.
    \item RQ2: Are there any factors which decides the success of RAG as  \emph{Virtual Teaching Assistant} or \emph{Teaching Aid}.
\end{itemize}

We recruited five faculty members in the Computer Science Department at The University of Texas at Tyler, USA, across Junior (3000), Senior (4000), and Graduate (5000, 6000) level courses. We developed individual RAG models, as illustrated in Section \ref{RAG}, for faculty members, and designed two tasks to investigate the potential of RAG for our research questions:
\begin{itemize}
    \item Task A: Can RAG aid in generation of questions for RQ1.
    \item Task B: Can RAG aid in answering questions of students RQ2.
\end{itemize}

The faculty members were allowed to probe the respective RAG systems for any tasks they deemed fit and specifically for the above two tasks. The investigation period were followed by a questionnaire, as illustrated in Section \ref{Q} to obtain ratings and opinions on the challenges, limitations, and potential improvements required for the use of RAG as faculty and student aid for our two tasks. The questionnaire allowed the faculty to rate the tasks on a Likert scale for RQ1 and RQ2. While feedback for RQ1 and RQ2 were obtained through direct questioning, the answer to RQ3 was interpreted from the difference in ratings for Task A and B between the faculty members at different class levels and we discuss it further in Section \ref{discussions}. 

\subsection{RAG Model}
\label{RAG}

By leveraging the vast amounts of structured and unstructured data available, RAG offers the potential to improve the quality and accuracy of language model outputs, bridging the gap between language generation and real-world knowledge. We use the open-source Gemini LLM model \cite{team2023gemini} to develop our RAG system, which is a cutting-edge large language model developed with an emphasis on adaptability, efficiency, and performance, influencing our choice over other LLM models. Moreover, Gemini was most particularly selected for our investigation owing to its reported prowess in solving mathematical problems \cite{ahmed2024gemini} along with other NLP tasks, which is an important task for educational settings, and other studies on Question Answering tasks yielding significantly high performance \cite{pal2024gemini} with the added advantage of being free, unlike outperforming models such as GPT-4.0, which is paid for API calls. Each faculty member in this study provided a set of slides or materials used in classroom teaching to develop their personalized RAG system. The slides contain text-based content, formulas, algorithms, code, graphs, and statistical data.

\begin{figure*}
     \centering
     \begin{subfigure}{0.45\textwidth}
        \includegraphics[width=\linewidth]{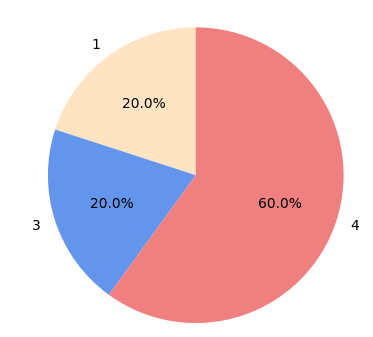}
    \end{subfigure}
    \hfill
     \begin{subfigure}{0.45\textwidth}
        \includegraphics[width=\linewidth]{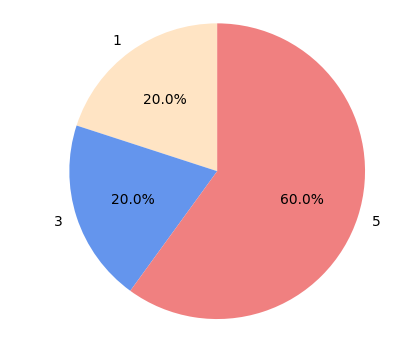}
    \end{subfigure}
    \hfill
     \begin{subfigure}{0.45\textwidth}
        \includegraphics[width=\linewidth]{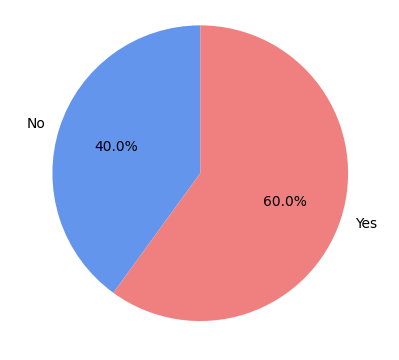}
    \end{subfigure}
    \hfill
     \begin{subfigure}{0.45\textwidth}
        \includegraphics[width=\linewidth]{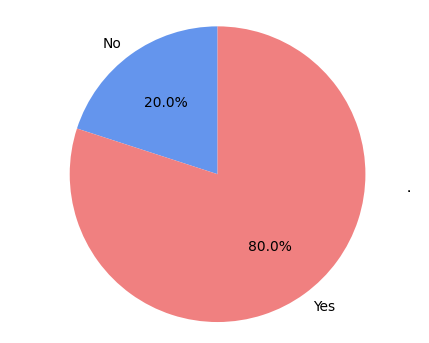}
    \end{subfigure}
     \hfill
        \caption{Faculty Feedback Rating on 5-Point Likert Scale for 4 questions obtained through Questionnaire: Top Left: Can LLMs aid generate assignment questions?, Top Right: Can LLMs aid in answering questions of students ?, Bottom left: Can LLMs be used as teaching assitant ?,and Bottom Right: Can LLMs be used as teaching aid for faculty in computer science classes ?}
        \label{fig:rating}
\end{figure*}

\subsection{Questionnaire}
\label{Q}
In this section, we discuss our questionnaire, which was presented to faculty members following their usage of the RAG system. This study incorporated factors such as perceived ease of use, perceived usefulness, attitudes, and intention to use, following the TAM model \cite{davis1989perceived}. The questionnaire contained three types of questions. For tasks A and B, as introduced in Section \ref{design}, the items were assessed using a 5-point Likert scale, with scores ranging from 1 to 5 as follows: strongly disagree = 1, disagree = 2, neutral = 3, agree = 4, and strongly agree = 5. Suggestions for improvements required for feasible use in daily teaching assistance, as teaching assistants, and general teaching aids were obtained through written answers to understand the limitations and potential of RAG LLMs. We also present two \emph{Yes} and \emph{No} answer questions seeking votes on the potential of RAG LLM as a teaching aid and teaching assistants in computer science higher education classes. No fixed period for usage of the RAG system was defined, and faculty members participating populated the questionnaire on their own timeline once they were satisfied with their investigation.

\section{Faculty Perspective: Potential and Limitation Discussions}
\label{perspectives}

In this section, we present the ratings and feedback obtained from the faculty members and are presented in four subsections of Task ratings (Section \ref{taskR}), RQ1 and RQ2 rating (Section \ref{RQ12R}), Faculty Feedback (Section \ref{FP}), and lastly RQ3 Interpretation (Section \ref{RQ3I}).

\subsection{Task Ratings}
\label{taskR}
\begin{table}[htbp]
\caption{Faculty opinion on Task A and Task B by different course level. A \checkmark represents a vote in the ability of the RAG and a X denotes a no-confidence on the task.}
\begin{center}
\begin{tabular}{|c|c|c|c|}
\hline
\textbf{Faculty} & \textbf{\textit{Course Level}}& \textbf{\textit{Task A}}& \textbf{\textit{Task B}} \\
\hline
Faculty 1 & 3000 & \checkmark & \checkmark\\
\hline
Faculty 2 & 4000 & \checkmark & \checkmark\\
\hline
Faculty 3 & 5000 & X & \checkmark\\
\hline
Faculty 4 & 5000 & X & \checkmark\\
\hline
Faculty 5 & 3000 & \checkmark & \checkmark\\
\hline
\end{tabular}
\label{baseline}
\end{center}
\end{table}

Our investigation and survey were conducted with five faculty members, allowing for the development of individualized RAG models, as well as a deeper dive into understanding the potential and challenges of RAG in higher educational settings, as illustrated in Section \ref{methodology}. Our five-member faculty cohort study results demonstrated an $80\%$ acceptance rate for Task A and $100\%$ acceptance rate for Task B, with any rating over three counted towards acceptance and below three to be rejected. The key ratings are shown in Fig. \ref{fig:rating} and the feedback can be illustrated as follows for our two tasks:
\begin{itemize}
    \item Task A: Three faculty members found it useful for task A of generating of assignment questions with a rating of four out of five, one faculty member rated the system one  (lowest rating score) and another faculty member provided a rating of three as presented in Fig. \ref{fig:rating} (Top Left).
    \item Task B: Three faculty members found RAG models to be helpful for task B of answering student questions and answers, and provided a high rating of five (highest rating score) out of five possible points and one faculty member rated it a 3 (neutral). The faculty members who did not find it useful rated it again, as shown in Fig. \ref{fig:rating} (Top Right).
\end{itemize}

\subsection{RQ 1 and RQ 2 Rating}
\label{RQ12R}
In this subsection, based on the ratings provided by the faculty, we present consolidated RQ1 and RQ2 (Research Questions). Faculty ratings are illustrated as follows:
\begin{itemize}
    \item RQ1: A $60\%$ agreement was recorded in favor of using RAG as a teaching aid for faculty in computer science higher education. Two faculty members did not find a system capable of undertaking Task A and disagreed with this potential application, as illustrated in Fig. \ref{fig:rating} (Bottom Left).
    \item RQ2: A $80\%$ agreement was reached by all cohort faculty members in the potential of our RAG system as a virtual teaching assistant for computer science higher education as illustrated in Fig. \ref{fig:rating} (Bottom Right).
\end{itemize}  

A majority agreement allows us to see the potential of the retrieval augmented generation technique using state-of-the-art LLMs in higher-educational settings. There seems to be better acceptance owing to higher potential as virtual \emph{Teaching Assistant} over \emph{Teaching Aid}. The feedback and opinions reported in Section \ref{FP} allow us to understand the improvements that would allow for full-scale deployment and acceptance with improved ratings.

\subsection{Faculty Perspectives}
\label{FP}
To delve into reasoning for rating and understanding the potential and challenges of RAG systems in computer science higher education, we investigate the reported feedback on the same. The feedback directly provides information on the improvements needed for large-scale deployment and implementation, with improved faculty ratings. The feedback from each of the cohort study members was filtered and presented to maintain relevance to the two tasks and can be summarized as follows:
\begin{itemize}
    \item Faculty 1: The RAG system can help answer student's questions when trying to complete homework assignments but would benefit from multiple sources of information such as multiple instructor slides and multiple textbooks. Furthermore, it needs a mechanism for reporting so that faculty can check usage/correctness of answers provided to students, and needs a mechanism for the interface so that students can email bot's responses to faculty members if they feel that they are being given incorrect information.
    \item Faculty 2: While the systems can help students as a Teaching Assistant to some extent, it needs better capabilities to solve simple problems based on equations that were discussed in the class.
    \item Faculty 3: The system can answer basic questions and would benefit from integrating general knowledge. The faculty members also mentioned that they would not use this bot in this form in their courses.
    \item Faculty 4: No, it still needs improvement and should be trained with more variations of prompts. For now, it is only text-based; it needs to be compatible with non-text files.
    \item Faculty 5: The system have capability of answering questions for students and generating basic question answers as quizlets for students. However, the model needs the context of different types of questions, such as MCQs and short answers, to generate questions or assignments. Moreover, the model does not perform well on the information from the images.
\end{itemize}

\subsection{RQ3 Interpretation}
\label{RQ3I} 

In this section, we analyze the interpretation of RQ3 based on faculty feedback and ratings of RQ1 and RQ2, as discussed in the previous sections. Faculty members 3 and 4, both teaching graduate-level classes, reported a lack of confidence in Task A, as shown in Table \ref{baseline}. Their feedback indicates that improvements are needed through training on mixed data, with the primary limitation being reliance on text-format files. This limitation was further corroborated by Faculty 5, who highlighted its inadequacy in inferring knowledge from images. Additionally, Faculty 2 suggested the need for enhanced capabilities to solve simple problems based on equations. Collectively, these feedback points suggest that direct application of RAG is more suitable for theoretical classes that do not extensively use images and equations. While we do not quantify performance, our focus is on reporting faculty views and understanding the potential and limitations of RAG in a higher education setting.

\section{Discussion and Future Work}
\label{discussions}
From the collected ratings and feedback, it is evident that while RAG demonstrates significant potential as both a \textit{Teaching Assistant} and \textit{Teaching Aid}, there are specific suggestions that need to be addressed for effective deployment and widespread integration in higher education. One prominent recommendation is the adoption of an expert-in-the-loop approach, which allows faculty members to monitor usage and verify the correctness of the model's responses, as highlighted by Faculty 1. Additionally, integrating multiple sources of knowledge beyond class slides, such as textbooks and reference documents, could enhance the system's utility. Another critical area of improvement is the capacity of the model to infer knowledge from images and equations.

For the specific task of serving as a \textit{teaching aid}, the models should be fine-tuned to differentiate between various types of questions, including multiple-choice questions (MCQs) and, short-answer, long-answer, and fill-in-the-blank questions. Our results from RQ3 indicate that the current limitations in inferring equations and their knowledge suggest a better potential for RAG in theoretical classes.

In future work, we plan to incorporate these suggestions and conduct large-scale, IRB-approved studies involving students from the computer science department and school of medicine. This will help us to further refine the system and evaluate its effectiveness in real-world educational settings.

\section{Conclusion}
Recent transformative developments in Deep Learning (DL) and Large Language Models (LLMs) have led to their integration in the realm of education. LLMs, with their advanced capabilities in natural language processing, have further amplified these prospects by enabling sophisticated text generation, comprehension, and interpretation tasks. This study specifically explored the application of Retrieval Augmented Generation (RAG), an emerging paradigm that combines the strengths of LLMs with external knowledge sources, thereby enhancing models' ability to generate precise and informative responses. To the best of the author's knowledge, this is the first work on the investigation of faculty perspectives for specialized RAG models tailored for computer science higher education as \emph{Virtual Teaching Assistant} and \emph{Teaching Aid}. We developed LLM-based RAG systems for use in junior, senior, and graduate-level computer science classes, and recruited faculty members to assess the potential of the systems. We identified the limitations and areas for improvement in integrating RAG systems into digital education, with the aim of facilitating their large-scale deployment and effectiveness. The feedback and ratings collected from faculty members highlighted both the promise and current shortcomings of RAG models, suggesting areas for further improvement to enhance their effectiveness and reliability in educational applications. Through our detailed discussion of faculty perspectives and the identification of necessary improvements, we hope to accelerate progress towards a more effective and widely accepted use of these technologies in higher education.

\bibliographystyle{ACM-Reference-Format}
\bibliography{sample-manuscript}

\end{document}